\documentstyle[12pt]{article}
\begin{document}

\begin{center}

{\bf Some problems of the pQCD jet calculus}\\

\vspace{2mm}

I.M. Dremin\\

Lebedev Physical Institute, Moscow\\

\end{center}

\begin{abstract}

Some problems of the perturbative quantum chromodynamics (pQCD) jet calculus
are discussed. The first one is related
to the terminology of the order of calculation. Due to cancelation of LO and NLO
terms in the ratio of mean multiplicities in gluon and quark jets $r$ the
nowadays obtained results about it should be called as 4NLO approximation.
The second problem reveals itself in calculations where corrections to some
values (in particular, to $r'$) are larger at present energies
than lower order terms. Some characteristics which do not suffer from this
deficiency are proposed. Next problem lies in interpretation of the negative
values of cumulant moments which are considered as an indication to the
replacement of attraction by repulsion in sets with definite particle contents.
Finally, the problem of the generalization of QCD equations for generating
functions is briefly discussed.

\end{abstract}

The numerous achievements of pQCD in prediction and description of properties
of quark and gluon jets are well known and described in many review papers
(see, e.g., \cite{dre1, koch, dgar, kowo, dr02}). Here, I would like to 
discuss some problems, related to these calculations and, often, left behind
the scene.

First, let me remind some simplest definitions \cite{5, dre1} concerning jet
multiplicities in QCD. The generating function $G$ is defined by the formula
\begin{equation}
G(y,u) = \sum_{n=0}^{\infty }P_{n}(y)u^{n}  ,                     \label{3}
\end{equation}
where $P_{n}(y)$ is the multiplicity distribution at the scale
$y=\ln (p\Theta /Q_0 )=\ln (2Q/Q_{0})$, $p$ is the initial momentum, $\Theta $
is the angle of the divergence of the jet (jet opening angle), assumed here to
be fixed, $Q$ is the jet virtuality,  $Q_{0}=$ const, $u$ is an auxiliary
variable.

The moments of the distribution are defined as
\begin{equation}
F_{q} = \frac {\sum_{n} P_{n}n(n-1)...(n-q+1)}{(\sum_{n} P_{n}n)^{q}} =
\frac {1}{\langle n \rangle ^{q}}\cdot \frac {d^{q}G(y,u)}{du^{q}}\vline _{u=1}, 
\label{4}
\end{equation}
\begin{equation}
K_{q} = \frac {1}{\langle n \rangle ^{q}}\cdot \frac {d^{q}\ln G(y,u)}{du^{q}}
\vline _{u=1}. \label{5}
\end{equation}
Here, $F_q$ are the factorial moments, and $K_q$ are the cumulant moments, 
responsible for total and genuine (irreducible to lower ranks) correlations,
correspondingly. These moments are not independent. They are connected by
definite relations which can easily be derived
from moments definitions in terms of the generating function:
\begin{equation}
F_{q} = \sum _{m=0}^{q-1} C_{q-1}^{m} K_{q-m} F_{m} .              \label{11}
\end{equation}

The QCD equations for the generating functions are:
\begin{eqnarray}
&G_{G}^{\prime }&= \int_{0}^{1}dxK_{G}^{G}(x)\gamma _{0}^{2}[G_{G}(y+\ln x)G_{G}
(y+\ln (1-x)) - G_{G}(y)] \nonumber \\ 
&+&n_{f}\int _{0}^{1}dxK_{G}^{F}(x)\gamma _{0}^{2}
[G_{F}(y+\ln x)G_{F}(y+\ln (1-x)) - G_{G}(y)] ,   \label{50}
\end{eqnarray}
\begin{equation}
G_{F}^{\prime } = \int _{0}^{1}dxK_{F}^{G}(x)\gamma _{0}^{2}[G_{G}(y+\ln x)
G_{F}(y+\ln (1-x)) - G_{F}(y)] ,                                   \label{51}
\end{equation}
where $G^{\prime }(y)=dG/dy ,$ $ n_f$ is the number of active flavours,
\begin{equation}
\gamma _{0}^{2} =\frac {2N_{c}\alpha _S}{\pi } ,                \label{52}
\end{equation}
the running coupling constant in the two-loop approximation is
\begin{equation}
\alpha _{S}(y)=\frac {2\pi }{\beta _{0}y}\left( 1-\frac {\beta _1}
{\beta _{0}^{2}}\cdot \frac {\ln 2y}{y}\right)+O(y^{-3}), \label{al}
\end{equation}
where
\begin{equation}
 \beta _{0}=\frac {11N_{c}-2n_f}{3}, \;\;\;\;\;\;
 \beta _1 =\frac {17N_c^2-n_f(5N_c+3C_F)}{3},
 \label{be}
\end{equation}
the labels $G$ and $F$ correspond to gluons and quarks,
and the kernels of the equations are
\begin{equation}
K_{G}^{G}(x) = \frac {1}{x} - (1-x)[2-x(1-x)] ,    \label{53}
\end{equation}
\begin{equation}
K_{G}^{F}(x) = \frac {1}{4N_c}[x^{2}+(1-x)^{2}] ,  \label{54}
\end{equation}
\begin{equation}
K_{F}^{G}(x) = \frac {C_F}{N_c}\left[ \frac {1}{x}-1+\frac {x}{2}\right] ,   
\label{55}
\end{equation}
$N_c$=3 is the number of colours, and $C_{F}=(N_{c}^{2}-1)/2N_{c}
=4/3$ in QCD.  

Herefrom, one can get equations for any moment of the multiplicity distribution
both in quark and gluon jets. One should just equate the terms with the same
powers of $u$ in both sides of the equations.
In particular, the equations for average  multiplicities read
\begin{eqnarray}
\langle n_G(y)\rangle ^{'} =\int dx\gamma _{0}^{2}[K_{G}^{G}(x)
(\langle n_G(y+\ln x)\rangle +\langle n_G(y+\ln (1-x)\rangle -\langle n_G(y)
\rangle ) \nonumber  \\
+n_{f}K_{G}^{F}(x)(\langle n_F(y+\ln x)\rangle +\langle n_F(y+
\ln (1-x)\rangle -\langle n_G(y)\rangle )],  \label{ng}
\end{eqnarray}
\begin{equation}
\langle n_F(y)\rangle ^{'} =\int dx\gamma _{0}^{2}K_{F}^{G}(x)
(\langle n_G(y+\ln x)\rangle +\langle n_F(y+\ln (1-x)\rangle -\langle n_F(y)
\rangle ).   \label{nq}
\end{equation}

Their solutions can be looked for as
\begin{equation}
\langle n_{G,F} \rangle \propto \exp (\int ^{y}\gamma _{G,F} (y\prime )dy\prime ).   \label{57}
\end{equation}
Using the perturbative expansion
\begin{equation}
\gamma _G \equiv \gamma = \gamma _{0}(1-a_{1}\gamma _{0}-a_{2}\gamma _{0}^{2}-a_3\gamma _0^3)+O(\gamma _{0}^{5})
 , \label{X}
\end{equation}
one gets the solution in the form \cite{5, dg, cdnt8}
\begin{equation}
\langle n_{G,F}\rangle=A_{G,F}y^{-a_{1}c^2 }\exp ( 2c\sqrt y+
\delta _{G,F}(y)),                                  \label{mul}
\end{equation}
where $c=(4N_c/\beta _0)^{1/2}$,
\begin{equation}
\delta _G(y)=\frac {c}{\sqrt y}[2a_2c^2+\frac {\beta _1}{\beta _{0}^{2}}
(\ln 2y+2)]+\frac {c^2}{y}[a_3c^2-\frac {a_1\beta _1}{\beta _{0}^{2}}
(\ln 2y +1)])+O(y^{-3/2}).
 \label{mean}
\end{equation}
Usually, in place of $\gamma _F$ the ratio of average multiplicities in gluon
and quark jets 
\begin{equation}
r=\frac {\langle n_G\rangle }{\langle n_F\rangle }=\frac {A_G}{A_F}
\exp(\delta _G(y)-\delta _F(y)) \label{rat}
\end{equation}
is introduced, and its perturbative expansion
\begin{equation}
r = r_0 (1-r_{1}\gamma _{0}-r_{2}\gamma _{0}^{2}-r_3\gamma _0^3)+
O(\gamma _{0}^{4})
  \label{Y}
\end{equation}
is used. The analytic expressions and numerical values of the parameters
$a_i, r_i$ for all $i\leq 3$ have been calculated from the perturbative
solutions of the above equations (the review is given in \cite{dgar}).
Within these approximations the experimental data about mean multiplicity
in $e^+e^-$-annihilation are well described as seen in Fig. 1 where
the notation $K\equiv 2A_F=2A_G/r_0$ is used.. However
the data about the ratio $r$ can be described with much lower accuracy
about 15$\%$ in such an analytical approach (see Fig. 2) even though each
subsequent perturbative approximation improves the agreement.

However, one should
mention here the quantitative fit provided by the computer solution of
the equation \cite{lo2, lo1}. This poses the question about the accuracy of
perturbative approximations for this particular characteristics and indicates
that the higher order corrections are still comparatively large for this
ratio up to the highest presently available energies. Let us also note that
the exact solutions of these equations for fixed
coupling constant were given in \cite{21, dhwa}.

The relation between the anomalous dimensions $\gamma $ of gluon and quark
jets is
\begin{equation}
\gamma _F=\gamma -\frac {r'}{r},  
\end{equation}
where
\begin{equation}
r'\equiv dr/dy=Br_0r_1\gamma _0^3[1+\frac {2r_2}{r_1}\gamma _0+
(\frac {3r_3}{r_1}+B_1)\gamma _0^2+O(\gamma _0^3)]       \label{derr}
\end{equation}
with $r_0=\frac {N_c}{C_F}=9/4; \; B=\beta _0/8N_c; \; B_1=\beta _1/4N_c\beta _0.$

Thus
\begin{equation}
\gamma _F = \gamma _{0}[1-a_{1}\gamma _{0}-(a_{2}+Br_1)\gamma _{0}^{2}-(a_3+
2Br_2+Br_1^2)\gamma _0^3-(a_4+B(3r_3+3r_2r_1+B_1r_1+r_1^3)) \gamma _{0}^{4}].
\label{gf}
\end{equation}
In these expressions we meet with two problems.

\begin{itemize}

\item{\bf Terminology.} 

The two leading terms in the energy behaviour of quark and gluon jet
multiplicities are absolutely the same as seen from Eq. (\ref{mul}) and cancel
in their ratio $r$ (\ref{rat}). Therefore this ratio is given by $r_0=9/4$ both
in the leading (LO) and next-to-leading (NLO) approximations. Thus, the common
notation DLA, which is used in Fig. 2 near the value $r$=9/4, should be
considered as LO+NLO-prediction of QCD for the ratio $r$. Therefore, the term
$r_1\gamma _0$ in (\ref{Y}) describes 2NLO corrections to the anomalous
dimension. However, in the literature, it is often called as a MLLA (NLO) term
what is wrong. Nevertheless, namely such notation is commonly used in Figures.
Here, in Fig. 2 we have used the notation with the letter $r$ added at the end.
It implies that, e.g., 3NLOr means that  the term with $\gamma _0^3$ in the
perturbative expansion of $r$ has been taken into account but it corresponds
to 4NLO-contribution to the anomalous dimension.

A misuse of the terminology for the anomalous dimensions $\gamma $'s and
for the ratio $r$ is clearly displayed in the explicit expression for
$\gamma _F$ (\ref{gf}). Its last 4NLO term contains $a_4$ which has not yet
been calculated. Together with it, the contribution from $r$ is present with
all terms calculated already and containing $r_i$ for $i\leq 3$ only. Thus,
let us stress again that in this sense one should say that "$r_3$"-term in $r$
corresponds to 4NLO contribution to the anomalous dimension of the quark jet
even though it is proportional to $\gamma _0^3$ in the perturbative expansion
of $r$.\\

\item{\bf Calculations.}

The cancellation of two leading terms in the ratio $r$ reveals itself also
in the proportionality of the scale (energy) derivative $r'$ to $\gamma _0^3$.
Therefore it can be calculated up to the terms $O(\gamma _0^5)$. The leading
term is very small (about 0.02 at the Z$^0$-resonance).
Asymptotically, all corrections vanish.
However, at present energies of Z$^0$, they are still quite
important. The second term in the brackets in (\ref{derr}) is larger than
1 since $2r_2/r_1 \approx 4.9$ and $\gamma \approx 0.45 - 0.5$. Even the 
third term is approximately about 0.4. The problem of convergence of the series
at Z$^0$-energies and below becomes crucial.
The derivative of the ratio $r$ (its energy slope) is very
sensitive to high order perturbative corrections.

Therefore, it is desirable to use at present energies such characteristics
which are less sensitive to these corrections. In particular,
these corrections partially cancel in the ratio of derivatives (slopes)
\begin{equation}
r^{(1)}=\frac {\langle n_G\rangle '}{\langle n_F\rangle '}. \label{ratd}
\end{equation}
The same is true for the ratio of curvatures (or second derivatives)
\begin{equation}
r^{(2)}=\frac {\langle n_G\rangle ''}{\langle n_F\rangle ''}. \label{radd}
\end{equation}
The QCD predictions for them
\begin{equation}
r<r^{(1)}<r^{(2)}<2.25
\end{equation}
were recently confirmed in experiment (see Figs. 3, 4 from \cite{488}). \\

\item{\bf Interpretation.}

Another question I'd like to raise concerns physical interpretation of
oscillations of cumulant moments in QCD which is not yet completely clarified.
Usually exploited phenomenological
distributions of the probability theory do not possess any oscillations.
E.g., all cumulant moments of the Poisson distribution are identically zero.
One interprets this as the absence of genuine correlations irreducible to
the lower-rank correlations. For the negative binomial distribution one
easily gets
\begin{equation}
H_q=\frac {K_q}{F_q}=\frac {2}{q(q+1)}>0.
\end{equation}
Since $F_q$ are always positive according to their definition, this inequality
implies the positive values of $K_q$.

In the leading order approximation, the gluodynamics equation for the 
generating function 
\begin{equation}
[\log G(y)]''=\gamma _0^2(G(y)-1)
\end{equation}
transforms in the relation
\begin{equation}
q^2K_q=F_q \;\;\;\; or \;\;\;\; H_q=\frac {1}{q^2}.   \label{hqq2}
\end{equation}
However already in the next-to-leading order $H_q$-moments become negative 
with a minimum at the rank $q_{min}\approx \frac {24}{11\gamma _0}+0.5\approx 5$
\cite{13}. This minimum is rather stable. It slowly moves to higher ranks with
energy increase and disappears in asymptotics as is required according to the
formula (\ref{hqq2}). At higher orders of the perturbative expansion, the
oscillations of higher rank cumulant moments
show up \cite{41}. They are confirmed in experiment \cite{dabg, sld} (see
Fig. 5).

Let me mention here that the plots of $D_q=q^2H_q$ instead of $H_q$ would be
even more instructive to reveal the oscillations. In this case they can be
easily compared to the LO prediction according to which $D_q^{LO}=1$. Also
the comparison to results of the negative binomial distribution would become
simplified. The plot of NBD results shows monotonic increase of $D_q^{NBD}$
from 1 at $q=1$ to 2 at $q\rightarrow \infty $ which is significantly different
from QCD oscillations.

Both the role of conservation laws and the changing character of the genuine
correlations can be blamed as originating these oscillations. If the latter
factor is important it would imply that attraction (clustering) is replaced
by repulsion (and vice versa) in particle systems with different number of
particles. It would be interesting to find other examples of such a behaviour
in hadronic systems.

\item{\bf Generalization.}

Finally, there exists the problem of possible generalization of the equations
for the generating functions. From one side,
we understand that even if treated as kinetic equations these
equations are limited by our ignorance of non-perturbative effects, simplified
treatment of conservation laws etc. Some phenomenological attempts to avoid
these limitations were attempted from the very beginning \cite{123, 127, 128}.
In \cite{123} it was proposed to treat hadronization of partons at the
final stage of jet evolution in analogy with the ionization in
electromagnetic cascades where it leads to their saturation and to the
finite length of the shower.
Three different stages of the cascade were considered in the modified kinetic
equations proposed in \cite{127, 128}. No
quantitative results were, however, obtained.

The most successful
modification of above equations was recently proposed \cite{eden} in
the framework of the dipole approach to QCD with more accurate kinematic
bounds. It has been shown that the ratio $r$ can be obtained in good agreement
with experimental data. Nevertheless, further study \cite{dede} of higher rank
moments of the multiplicity distribution predicted by the modified equations has
shown their extremely high sensitivity to higher orders of the perturbative
expansion. As shown in Fig. 6,
the moments diverge at high orders and the only trace of oscillations can be
noticed in the changing signs of the moments of the subsequent ranks. 
The results become inconclusive. Thus no successful generalization is at work
nowadays. Rather, the general trend shifted to the direct calculation of
non-perturbative effects in some jet characteristics (see, e.g., \cite{yud, bdmz}).

At the same time, the success of numerical solutions of the existing equations
\cite{lo2, lo1} raises the question if the generalization will give any other
noticable contribution and our failure to describe more precisely the ratio $r$
could be just due some defects of the purely perturbative expansion at available
energies. More rigorous treatment of the numerical solutions of the equations
should be done. Moreover, it was claimed recently \cite{hama} that the
renormalization group improvement of the perturbative results gives rise
to good description of experimental data.

\end{itemize}

In conclusion, I'd say that, even though some principal questions concerning
the calculation of some properties of quark-gluon jets and the validity
of QCD equations for the generating functions at higher orders are not yet
resolved, the practical accuracy of the pQCD calculations is high enough,
especially, in view of the rather large expansion parameter.\\

This work is supported by the RFBR grant 00-02-16101.\\

{\bf Figure captions.}\\

Fig. 1. The energy dependence of average multiplicity  of charged particles in 
$e^+e^-$-annihilation. The results of different fits according to formulas of
perturbative QCD and of the Monte Carlo models are shown ( the solid and dotted
lines are the fits of formula (\ref{mean}) with one and two adjusted parameters,
the dashed line is given by the HERWIG Monte Carlo model; the vertically shaded
area indicates the gluon jet data multiplied by the theoretical value of the
ratio $r$ (\ref{Y})).\\

Fig. 2. The experimentally measured ratio $r$ of multiplicities in gluon and quark
jets as a function of energy in comparison with the predictions of analytical QCD
and of the Monte Carlo model HERWIG (different QCD approximations, described
in this paper, as well as $r$($\epsilon $) with integration limits
$e^{-y}$ and 1-$e^{-y}$ in Eqns (\ref{50}), (\ref{51}) are indicated at
the corresponding lines).\\

Fig. 3. The ratio of the slopes of the energy dependences of mean
multiplicities in gluon and quark jets according to experimental data and some
theoretical calculations.\\

Fig. 4. The ratio of the curvatures of the energy dependences of mean
multiplicities in gluon and quark jets according to experimental data and some
theoretical calculations.\\

Fig. 5. The measured ratio $H_q$ of the cumulant and factorial moments oscillates
as a function of the rank $q$ according to experimental data on multiplicity
distributions of charged particles in $e^+e^-$-annihilation at the $Z^0$ energy
(the inset in the upper right corner shows the data for the moments of
the ranks 2, 3 and 4).\\

Fig. 6. The $H_q$-moments in the modified dipole approach \cite{eden, dede}
drastically diverge at higher orders for large ranks $q$ with changing the sign
at subsequent ranks.\\


\begin{thebibliography}{99}
\bibitem{dre1}
Dremin I M  {\it Phys.-Uspekhi} {\bf 37} 715 (1994)
\bibitem{koch}
Khoze V A and Ochs W {\it Int. J. Mod. Phys.} {\bf A 12} 2949 (1997) 
\bibitem{dgar}
Dremin I M and Gary J W {\it Phys. Rep.} {\bf 349} 301 (2001)
\bibitem{kowo}
Khoze V A, Ochs W and Wosiek J in {\it "Handbook of QCD" (Ioffe Festschrift)}
(WSPC, Singapore) (to be published); hep-ph/0009298
\bibitem{dr02}
Dremin I M {\it Phys.-Uspekhi} {\bf 47} N5 (2002)
\bibitem{5}
Dokshitzer Yu L, Khoze V A, Mueller A H and Troyan S I {\it Basics of
perturbative QCD} ed. J. Tran Thanh Van
(Gif-sur-Yvette, Editions Frontieres, 1991).
\bibitem{dg}
Dremin I M and Gary J W {\it Phys. Lett.} {\bf B 459} 341 (1999) 
\bibitem{cdnt8}
Capella A, Dremin I M, Gary J W, Nechitailo V A and Tran Thanh Van J
{\it Phys. Rev.} {\bf D 61} 074009 (2000) 
\bibitem{lo2}
Lupia S {\it Phys. Lett.} {\bf B 439} 150 (1998);
{\it Proc. XXXIII Moriond conf. "QCD and strong interactions"}
March 1998, ed. J. Tran Thanh Van (Editions Frontieres, Gif-sur-Yvette, 1998)
p. 363
\bibitem{lo1}
Lupia S and Ochs W {\it Phys. Lett.} {\bf B 418} 214 (1998); {\it Nucl. Phys.
(Proc. Suppl.)} {\bf B 64} 74 (1998)
\bibitem{21}
Dremin I M and Hwa R C {\it Phys. Rev.} {\bf D 49} 5805 (1994)
\bibitem{dhwa}
Dremin I M and Hwa R C {\it Phys. Lett.} {\bf B 324} 477 (1994)
\bibitem{488}
Gary J W in {\it Proc. 31 Int. Symp. on Multiparticle Dynamics} 1-7 Sept. 2001,
Datong, China, Eds. Liu L., Wu Y., World Scientific, Singapore, 2002
(to be published); \\
OPAL Collaboration Physics Note PN488, 2001 
\bibitem{13}
Dremin I M {\it Phys. Lett.} {\bf B 313} 209 (1993)   
\bibitem{41}
Dremin I M and Nechitailo V A {\it Mod. Phys. Lett.} {\bf A 9} 1471 (1994);
{\it JETP Lett.} {\bf 58} 881 (1993)
\bibitem{dabg}
Dremin I M, Arena V, Boca G et al {\it Phys. Lett.} {\bf B 336} 119 (1994) 
\bibitem{sld}
SLD Collaboration, Abe K et al {\it Phys. Lett.} {\bf B 371} 149 (1996)
\bibitem{123}
Dremin I M {\it Pis'ma v ZhETF} {\bf 31} 215 (1980); {\it JETP Lett.} {\bf 31}
185 (1980)
\bibitem{127}
Ellis J and Geiger K {\it Phys. Rev.} {\bf D 52} 1500 (1995)
\bibitem{128}
Ellis J and Geiger K {\it Nucl. Phys.} {\bf A 590} 609c (1995)
\bibitem{eden}
Eden P {\it Proc. XXXIV Moriond conf. "QCD and strong interactions"} March 1999,
 ed. J. Tran Thanh Van (Editions Frontieres, Gif-sur-Yvette, 1999)
\bibitem{dede}
Dremin I M and Eden P in {\it Proc. 31 Int. Symp. on Multiparticle Dynamics}
1-7 Sept. 2001, Datong, China, Eds. Liu L., Wu Y., World Scientific, Singapore, 2002
(to be published)
\bibitem{yud}
Dokshitzer Yu L Talk at this workshop, Durham, Dec. 2001.
\bibitem{bdmz}
Banfi A, Dokshitzer Yu L, Marchesini G and Zanderighi G {\it JHEP} {\bf 0007}:002
(2000); {\it Phys. Lett.} {\bf B508} 269 (2001); {\it JHEP} {\bf 0103}:007 (2001)
\bibitem{hama}
Hamacher K Talk at this workshop, Durham, Dec. 2001.

\end{thebibliography}
\end{document}